\documentclass{aa}  
\usepackage{graphicx}
\usepackage{txfonts}
\usepackage{lipsum}
\usepackage{subcaption}
\usepackage{lscape}
\usepackage{placeins}
\usepackage{wrapfig}
\usepackage{hyperref}

\titlerunning{A Cosmic Web of Satellites}
\authorrunning{S. Tavasoli et al.}

\begin{document}

\title{Cosmic environment as the primary driver of dwarf satellite statistics}

\author{S. Tavasoli\inst{1} \fnmsep \thanks{Corresponding Author: stavasoli@khu.ac.ir},
        \and P. Ghafour\inst{1} \fnmsep \inst{2}}
\institute{Department of Astronomy and High Energy Physics, Faculty of physics, Kharazmi University, Tehran 15719-14911, Iran 
\and
Department of Physics, Shahid Beheshti University, Tehran 1983969411, Iran }

\date{}

\abstract
{
    Context. Satellite dwarf galaxies provide key constraints on galaxy formation and evolution, since their abundance and spatial distribution reflect both the host properties and the large-scale environment.
  
    Aims. This study quantifies the dependence of satellite populations on the host stellar mass, morphology, and star formation activity across different environments, and traces their evolution with cosmic time within the $\Lambda$CDM framework.
  
    Methods. The Millennium-II simulation combined with the G11 semi-analytic model is used to construct consistent samples of host galaxies brighter than $M_{r}<-16$ and their satellites ($M_{\ast}\geq 3\times10^{5}\,M_{\odot}$, $M_{r}<-9$) within the virial radius. Satellite abundance and radial profiles are analysed in cluster, group, and void environments, and their evolution is traced from $z=2$ to $z=0$ across three host stellar mass bins.
  
    Results. Satellite abundance is correlated strongly with host stellar and bulge mass, whereas host morphology has little independent effect once stellar mass is accounted for. Dense environments suppress satellite populations relative to voids. Correlations between satellite abundance, specific star formation rate, and disk scale length become evident only in groups and clusters. At $z=0$, radial profiles show strong central concentrations in voids, flattened distributions in clusters, and intermediate trends in groups. Their redshift evolution reveals progressive flattening for low- and intermediate-mass hosts in dense environments, stability for massive hosts, and increasing central concentration in voids. The cosmic evolution of satellite abundance further highlights distinct pathways: gradual accumulation in voids, mass-dependent trends in groups, and strong late-time suppression in clusters.
  
    Conclusions. The distribution and abundance of satellite galaxies are governed primarily by environment, with the host stellar mass and cosmic epoch acting as secondary modulators. From the dense interiors of clusters to the isolation of voids, large-scale structure imprints systematic signatures on satellite system assembly. Forthcoming wide-field surveys such as the Legacy Survey of Space and Time (LSST), the ESA Euclid mission (Euclid), and the Nancy Grace Roman Space Telescope are expected to provide stringent tests of these predictions and advance our understanding of the interplay between host properties, environment, and satellite evolution across cosmic time.
}

\maketitle
\nolinenumbers

\section{Introduction \label{sec:intro}}

Dwarf galaxies are the smallest and most abundant building blocks of the universe and rank among the most dark matter-dominated systems known (\cite{salucci2019distribution}). Their abundance, spatial distribution, kinematics, and internal dynamics make them powerful probes for testing cosmological models (e.g. \cite{moore1999dark,boylan2011too,kroupa2010local,ibata2014velocity,van2018galaxy,muller2019dwarf}). Consequently, studying the properties of dwarf galaxies is essential for advancing our understanding of galaxy formation and evolution.

However, due to their low luminosities and surface brightnesses, dwarf galaxies are challenging to detect at large distances (\cite{martin2019formation}). Consequently, our current understanding is predominantly based on observations within the Local Group (LG), where the majority of known dwarfs are satellites of the Milky Way (MW) and the Andromeda galaxy (M31) (\cite{metz2007spatial}). The population, intrinsic properties, and spatial distribution of LG satellites have provided a critical testing ground for galaxy formation models, revealing several challenges to the prevailing 
$\Lambda$ cold dark matter ($\Lambda$CDM) cosmological paradigm. These challenges include the missing satellites problem (\cite{klypin1999missing}), the too-big-to-fail problem (\cite{boylan2011too}), and the planes of satellites problem (\cite{pawlowski2013rotationally,muller2018whirling}).

While the $\Lambda$CDM model successfully explains the formation of large-scale structures (\cite{eisenstein2005detection}), its predictions on galactic and sub-galactic scales remain under ongoing evaluation (\cite{bullock2017small}). A more recent challenge has arisen from the discovery of empirical correlations that appear inconsistent with a purely dark matter-driven hierarchy. Notably, correlations have been found between the number of satellites ($N_{\mathrm{sat}}$) and the bulge mass ($M_{\mathrm{bulge}}$) and the bulge-to-total mass ratio ($M_{\mathrm{bulge}}/M_{*}$) of their host galaxies (\cite{lopez2016number,javanmardi2019number,javanmardi2020correlation}). Such correlations are unexpected within the $\Lambda$CDM framework, where the satellite population should correlate primarily with the host’s dark matter halo mass and, by extension, its total dynamical mass or rotation velocity, not with the baryonic sub-structure (\cite{kravtsov2010dark}).

In the $\Lambda$CDM framework, galaxies form through the condensation of baryons at the centres of dark matter halos (\cite{white1978core}). Major galaxies grow hierarchically via the mergers of earlier dwarf-sized progenitors (\cite{cole2000hierarchical}). Accordingly, the majority of dwarf satellites are predicted to be primordial, dark matter-rich objects, with a minority population comprised of younger, tidal debris-born systems that may be significantly deficient in dark matter (e.g. \cite{mcconnachie2012observed}).

Recent observational campaigns have substantially expanded the census of dwarf satellites surrounding galaxies beyond the Local Group (LG) (e.g. \cite{gulzow2024stellar}). These new data offer valuable opportunities to investigate the satellite population, its dependence on host galaxy properties, and the influence of the large-scale environment. To date, the role of environment in shaping the number of dwarf satellites and their correlation with the host properties remains one of the least explored facets of these phenomena.

The potential relationship between the properties of galaxies in different large-scale environments, such as voids, groups, and clusters, and their dwarf satellite populations has not been quantitatively investigated within the standard cosmological model. In this work, this question is addressed using data from the Millennium-II simulation (\cite{boylan2009resolving}; hereafter MS-II), combined with the semi-analytic galaxy formation model of \cite{guo2013galaxy} (hereafter G11). The MS-II simulation provides sufficient resolution for a detailed study of dwarf satellite galaxies. For example, it has been successfully employed in previous statistical analyses of satellite populations around Andromeda-like galaxies within the $\Lambda$CDM framework, such as those conducted by \cite{bahl2014comparison} and \cite{javanmardi2019number}. This analysis aims to provide new observational constraints on galaxy formation models by testing whether the satellite population is shaped primarily by the host's dark matter halo, as predicted by $\Lambda$CDM, or whether baryonic processes and environmental history play a more significant role.

This paper is organized as follows. Section~\ref{sec:method} describes the dataset used in this study, the cosmic‑web classification procedure, and the identification of satellite galaxies across different environments. The results and their analysis are presented in Section~\ref{sec:result}. Finally, Section~\ref{sec:sum} provides a summary of the main findings and the concluding remarks.

\section{Method \label{sec:method}}
\subsection{Millennium-II simulation} \label{sec:Mill}

To study the effect of environment on dwarf satellite galaxies, the sample of host halos and their satellite populations is selected from the semi-analytic model of galaxy formation developed by \cite{guo2011dwarf}, applied to the MS-II simulation ( \cite{boylan2009resolving}).

In the G11 SAM model, baryonic matter is assigned to dark matter halos according to the cosmological baryon fraction derived from the first--year WMAP results (\cite{spergel2003first}). The effective baryon fraction that collapses into halos depends on the halo mass and the redshift and is described using the fitting function proposed by \cite{Gnedin2000Reionization}, with the redshift-dependent characteristic halo mass taken from \cite{Okamoto2008Massloss}. This prescription accounts for the suppression of baryon accretion in low-mass halos caused by photoheating from the ultraviolet background after reionization (\cite{Doroshkevich1967Origin,Efstathiou1992Photoionization}).

Within each halo, the baryonic matter is distributed among several components: a hot gas halo, a cold gas disk, a stellar disk, a stellar bulge, and an ejecta reservoir (\cite{Springel2005Millennium,Croton2006SAM}). The baryons that do accrete into halos are initially assumed to form a diffuse hot gas atmosphere (\cite{Dekel2009ColdFlows}), and in this model, satellite galaxies are also allowed to retain their own hot gas halos, which can be dynamically stripped by tidal and ram-pressure effects, although this gas may continue to cool onto the satellite galaxy and supply additional fuel for star formation. Gas cools radiatively from the hot halo and settles into a rotationally supported cold gas disk, where stars form according to a simplified version of the empirical relation proposed by \cite{Kennicutt1998SchmidtLaw}, building the stellar disk. The rate at which gas accretes onto the central galaxy depends on the gas cooling time (\cite{Springel2001Subhalos}) and the halo dynamical time (\cite{DeLucia2004ChemicalEnrichment}). Feedback from star formation, particularly supernova feedback (\cite{Larson1974Collapse}), can reheat cold gas and return it to the hot halo or eject it into the ejecta reservoir, significantly influencing the evolution of low-mass galaxies and their metallicities. The bulge component grows through galaxy mergers and disk instabilities, with three main channels included in the model: major mergers, minor mergers, and disk buckling. Major mergers transfer all stellar material from the progenitors into a spheroidal remnant, whereas in minor mergers the disk of the larger galaxy remains intact while its bulge accretes the stars from the smaller progenitor. The continuous exchange of mass between the different baryonic reservoirs through cooling, star formation, feedback, mergers, and disk instabilities ultimately drives the structural evolution of the model galaxies.

Galaxy luminosities and magnitudes are computed from the star formation and chemical enrichment histories predicted for each galaxy. At each time-step, newly formed stellar populations are recorded with their corresponding ages and metallicities, and the integrated spectral energy distribution (SED) is constructed using stellar population synthesis models (\cite{BruzualCharlot2003}) together with a Chabrier initial mass function (IMF), which provides a reduced fraction of low-mass stars and better agreement with observational constraints (\cite{DeLucia2004ChemicalEnrichment}). Dust attenuation is then applied using a slab model that depends on the cold gas content, metallicity, and geometry of the galaxy (\cite{GuoWhite2009}). In addition, a redshift dependence is included, so that, for galaxies with a given gas metallicity, the gas-to-dust ratio is assumed to be higher at earlier cosmic times than in the local universe (\cite{Steidel2004LBGs,Quadri2008Clustering}). The resulting SEDs are used to compute galaxy magnitudes in various observational filter systems, enabling direct comparisons with observational surveys. Further details of these procedures are provided in Sections 2, 3, and 4 of \cite{guo2011dwarf}.

The MS-II simulation traces galaxy formation within a cubic volume of $100 \ h^{-1} \ Mpc$ on each side, using $2160^{3}$ dark matter particles with a mass resolution of $9.45 \times 10^{6} \ M_{\odot}$. It adopts a flat $\Lambda$CDM cosmology consistent with the WMAP1 data \cite{spergel2003first}, with parameters: $\Omega_{\Lambda} = 0.75$, $\Omega_{m} = 0.25$, $\Omega_{b} = 0.045$, $h = 0.73$, $n = 1$, and $\sigma_{8} = 0.9$.

Within MS-II, subhalos with masses as low as $2 \times 10^{8}$ $M_{\odot}$ can be resolved (\cite{boylan2009resolving}). As demonstrated by \cite{guo2011dwarf}, this resolution renders the simulation well suited for investigating dwarf satellite populations across diverse cosmic web environments.

\subsection{Environments} \label{sec:environment}

To investigate the role of a large-scale environment, the galaxy sample is classified into three categories based on local density: clusters, groups, and voids. The classification begins with the linking of dark matter halos. In MS-II, friends-of-friends (FoF) halos are linked by connecting dark matter particles separated by less than 0.2 times the mean inter-particle separation \cite{davis1985evolution}. Within these FoF halos, self-bound subhalos are identified using the \textsc{subfind} algorithm \cite{springel2001hydrodynamic}. In the semi-analytic model, each subhalo is associated with a galaxy whose dynamics is governed by the underlying dark matter structure \cite{guo2011dwarf}.

Cluster environments, defined as extremely high-density regions, correspond to FoF groups containing at least 100 galaxy members with $M_{r} < -16$ at $z=0$. This criterion yields 44 distinct galaxy clusters. Group environments, representing intermediate-density regions, are defined as FoF groups containing four to eight galaxy members (with $M_{r} < -16$ at $z=0$), resulting in 2206 galaxy groups.

Since the dominant physical mechanisms driving galaxy evolution in dense environments (e.g. ram-pressure stripping (\cite{boselli2022ram}), tidal interactions(\cite{hahn2009tidal})) operate most effectively within the deep gravitational potential wells of such systems (\cite{wetzel2012galaxy}), only galaxies residing within the virial radius of their host cluster or group are considered in the analysis. After applying this spatial constraint, the resulting samples comprise approximately $\sim 4000$ galaxies in clusters and $\sim 6500$ galaxies in groups.

For under-dense regions, the void catalogue of \cite{tavasoli2013challenge} is employed, constructed from MS-II using the method of \cite{aikio1998simple}. This algorithm identifies large under-dense regions in the cosmic web with effective radii greater than $7 \ Mpc$. From this catalogue, all galaxies residing in voids with $M_{r} < -16$ at $z=0$ are selected, yielding 2356 galaxies within 179 distinct cosmic voids. This methodology yields robust and statistically significant samples, enabling comparative analysis across the full spectrum of cosmic density environments.

\subsection{Dwarf satellite} \label{sec:satellite}

The integrity of the results relies on the rigorous and consistent identification of satellite galaxies across diverse cosmic environments, ranging from dense clusters to sparse voids. To enable robust comparative analysis, a unified methodology for catalogue construction is adopted, following the framework of \cite{javanmardi2019number}, who similarly used a Millennium-II-based galaxy catalogue to investigate satellite systems. The selection criteria used to construct a uniform and unbiased satellite catalogue across clusters, groups, and voids are as follows:
\begin{enumerate}[i)]
	\item All halos containing any galaxy with a stellar mass exceeding 0.8 times that of the central galaxy are excluded. This criterion ensures that the identified central galaxy serves as the dominant gravitational anchor of its host halo.
	
	\item Only satellite galaxies (i.e. those residing in subhalos) located within the virial radius ($R_{vir}$) of the central galaxy and exhibiting an $r$-band absolute magnitude of $M_{r} < -9$ are considered. These selection criteria are adopted from typical conditions employed in observational studies of classical satellites (e.g. \cite{martin2009pandas,carlsten2021structures}).
	
	\item Only satellites with stellar masses $M_{*} > 3 \times 10^{5}$ $M_{\odot}$ are retained, matching the luminosity cut of $LV>2\times10^{5}$ used by \cite{kroupa2010local} and assuming a mass-to-light ratio of 1.5 (\cite{dabringhausen2016extensive}).
\end{enumerate}
Thus, gravitationally bound satellite galaxies are systematically identified by analyzing their spatial distribution, kinematic properties, and dark matter halo membership relative to a central host galaxy. This procedure enables a consistent census of satellite populations in environments of varying density, thereby mitigating potential selection biases.

In the semi-analytic model, each satellite is defined as a galaxy hosted by a resolved subhalo, ensuring that the satellite classification remains physically motivated and consistent with the underlying dark matter structure.

Applying these criteria to the $z=0$ snapshot of the MS-II simulation yields approximately 
2000, 3600, and 2800 host galaxies in void, group, and cluster environments, respectively,
accompanied by a total of ~2400, 17200, and 5200 satellites. The substantial population statistics obtained serve as a reliable foundation for investigating correlations between satellite galaxy properties and their surrounding environmental context.

\section{Analysis and results 
	\label{sec:result}}
This section presents the principal findings of the investigation into the environmental dependence of satellite galaxy populations. The analysis explores how satellite abundance and spatial distribution vary with key host galaxy properties across clusters, groups, and voids. The results underscore the interplay between intrinsic host characteristics and external environmental factors in shaping satellite systems. The study is organized into two complementary components. First, statistical scaling relations between satellite abundance and host properties are established across distinct environments (\ref{sec:3.1}). Second, the spatial distribution of satellites is examined around hosts of varying masses and across multiple redshifts (\ref{sec:3.2}), and the assembly history of satellite systems is traced by following host galaxies backward in time, thereby offering a dynamic perspective on their evolution across cosmic epochs.

\subsection{Satellite abundance versus host properties 
	\label{sec:3.1}}
The analysis now turns to the dependence of satellite abundance on the intrinsic properties of host galaxies. Specifically, correlations are examined with stellar mass (\ref{sec:3.1.1}), bulge mass (\ref{sec:3.1.1}), morphological type (\ref{sec:3.1.2}), star formation activity (\ref{sec:3.1.3}), and disk scale length (\ref{sec:3.1.3}). These complementary diagnostics facilitate the assessment of the host characteristics that most strongly influence the richness of satellite systems.

\subsubsection{Dependence on stellar mass and bulge mass 
	\label{sec:3.1.1}}
\begin{figure}[h!]
	\centering
	\includegraphics[width=0.47\textwidth]{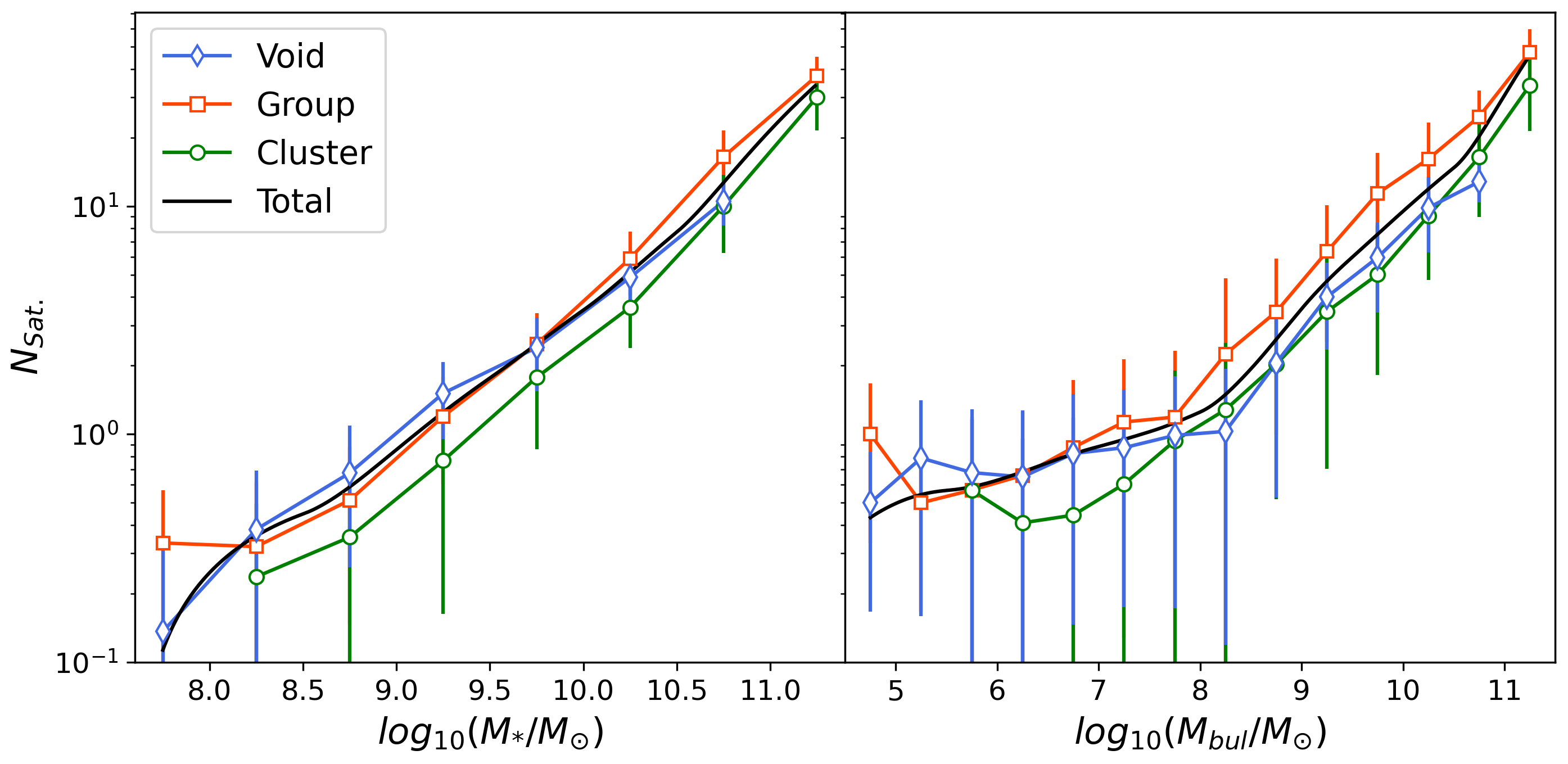}
	\caption{Mean trends in satellite number ($N_{sat.}$) and associated $1\sigma$ errors are shown as a function of stellar mass (left) and bulge mass (right). Different environments are indicated by colour: voids in blue, groups in orange, clusters in green, and the full population in black.}
	\label{Fig:1}
\end{figure}

The analysis begins with an examination of the relationship between satellite galaxy abundance and fundamental host mass properties. Figure \ref{Fig:1} presents the mean number of satellites as a function of total stellar mass $M_{*}$ (left panel) and bulge mass $M_{bul}$ (right panel) for host galaxies in voids (blue), groups (orange), clusters (green), and the full population (black).

The results reveal a strong linear correlation between satellite abundance and both host stellar mass and bulge mass across all environments. Satellite counts increase systematically from $\sim$0.3 for low-mass hosts ($\log_{10}(M_*/M_\odot) \sim 8$) to $\sim$40 for the most massive systems ($\log_{10}(M_*/M_\odot) \sim 11$). A similar trend is observed for the bulge mass, with satellite numbers increasing from $\sim$0.5 at $\log_{10}(M_{\text{bul}}/M_\odot) \sim 5$ to $\sim$40 at $\log_{10}(M_{\text{bul}}/M_\odot) \sim 11$. The linearity of these relations over nearly three orders of magnitude indicates that the processes governing satellite system formation and evolution scale uniformly with host mass, largely independent of environmental context.

Environmental trends remain consistent in both mass definitions. Cluster hosts (green) systematically contain fewer satellites at all stellar masses, whereas group hosts (orange) and void  hosts (blue) exhibit statistically similar abundances, as indicated by overlapping error bars. The persistent offset of cluster galaxies below the group and void relations suggests that dense environments via mechanisms such as ram-pressure stripping (\cite{boselli2022ram}), and satellite harassment (\cite{bialas2015occurrence}) systematically suppress satellite populations beyond what is expected from host mass alone. The parallel trends observed in groups and voids, despite their contrasting large scale densities, imply that satellite systems in these environments share similar formation histories and undergo relatively limited environmental disruption.

When considering the bulge mass, a distinct pattern emerges. Above $\log_{10}(M_{\text{bul}}/M_\odot) \sim 8$, group environments host significantly more satellites than both void and cluster environments. This enrichment becomes increasingly pronounced at higher bulge masses, indicating that groups provide particularly favourable conditions for sustaining satellite systems around bulge-dominated galaxies. By contrast, cluster hosts consistently show suppressed satellite populations across the entire bulge mass range, underscoring the efficiency of cluster-specific processes in limiting satellite abundance irrespective of bulge properties.

Clarification is warranted regarding the enhancement of satellite counts in group environments at high bulge masses ($M_{\text{bul}} > 10^8 M_{\odot}$). Given the natural correlation between total stellar mass and bulge mass (e.g. \cite{Mendez-Abreu2021}), one might suspect that this enhancement simply reflects the higher stellar masses of group galaxies. However, as shown in the left panel of Figure~\ref{Fig:1}, at fixed total stellar mass, satellite abundances in groups and voids are statistically indistinguishable across most of the mass range (only clusters show a systematic deficit). Even at the highest stellar masses ($M_* > 10^{10.5} M_{\odot}$), where the groups exhibit a marginal excess over the voids, the difference is far smaller than the enhancement observed at the fixed bulge mass. Therefore, the excess satellites in groups at fixed bulge mass cannot be explained by a simple stellar mass--bulge mass correlation. Instead, it points to a genuine environmental effect that specifically enhances satellite populations around bulge-dominated galaxies in group environments. Furthermore, our results are consistent with \citet{Muller2023}, who showed that at fixed stellar mass, early-type galaxies (which dominate the high bulge mass regime in groups) host richer satellite populations than late-type galaxies.

Error analysis shows substantially larger scatter for low bulge mass hosts across all environments, reflecting the stochastic nature of satellite system formation around galaxies with underdeveloped bulges. At higher bulge masses, the scatter decreases, indicating more stable and predictable satellite populations.

These results demonstrate that environmental effects on satellite systems manifest differently when considering bulge mass versus total stellar mass. The particularly rich satellite systems in group environments with massive bulges may reflect a balance between sufficient density to promote accretion and the absence of the destructive processes prevalent in clusters. Conversely, consistent suppression in clusters suggests that tidal stripping and harassment efficiently reduce satellite numbers regardless of bulge properties, while the larger scatter at low bulge masses points to more stochastic assembly histories.

\begin{figure}[h!]
	\centering
	\includegraphics[width=0.47\textwidth]{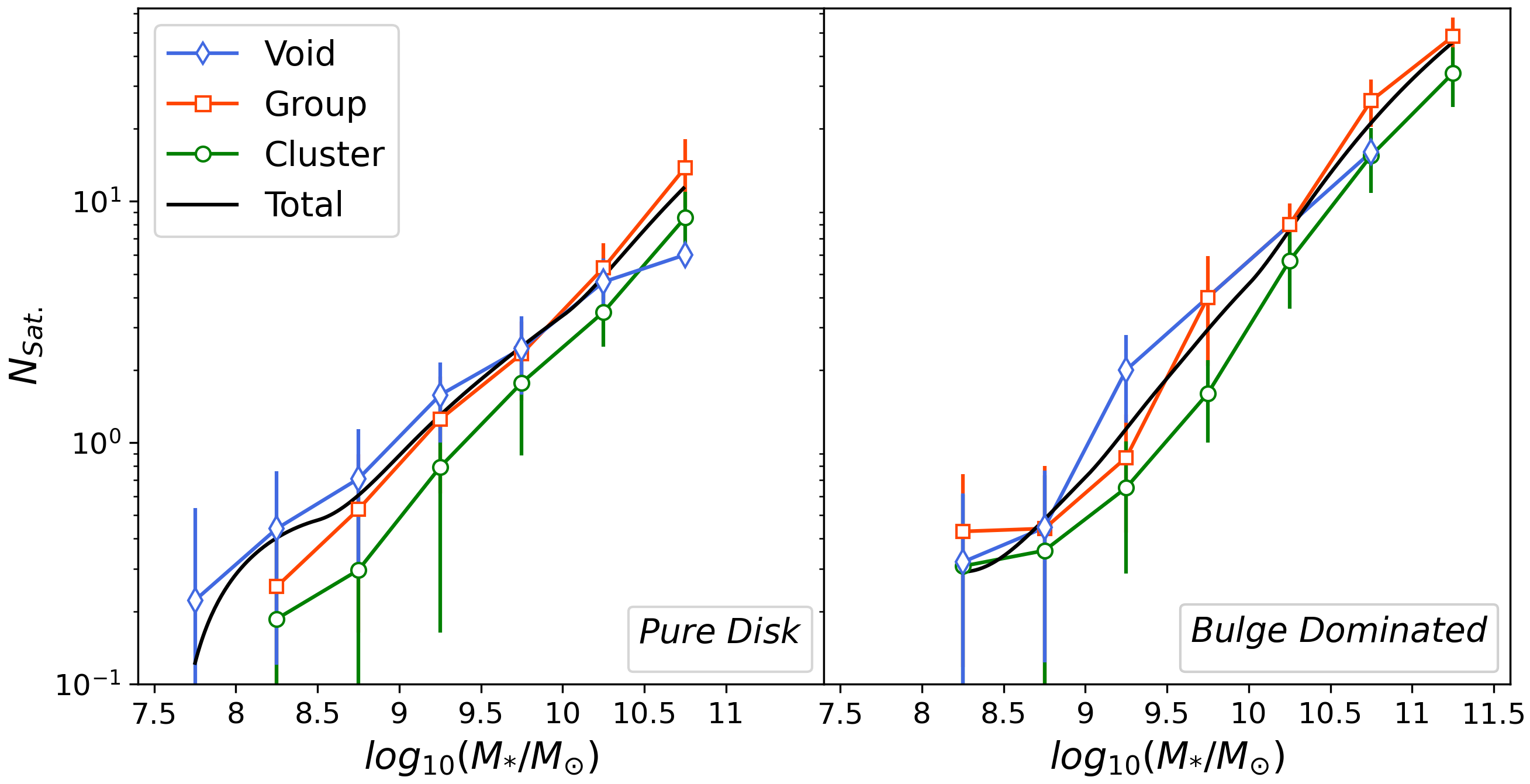}
	\caption{Mean trend of satellite number ($N_{sat.}$) is illustrated with respect to the stellar mass of the host galaxy for pure-disk (left) and bulge-dominant (right) galaxies. Various environments are shown in distinct colours.}
	\label{Fig:2}
\end{figure}

\subsubsection{Dependence on morphology 
	\label{sec:3.1.2}}
The connection between satellite abundance and host morphology has been extensively examined in previous studies (e.g. \cite{kroupa2010local,javanmardi2020correlation}), which report that galaxies with negligible bulges tend to host very few or no satellites, based on both observational data and simulations. The present analysis examines the satellite populations of negligible-bulge (pure-disk) and large-bulge (bulge-dominated) galaxies across cluster, group, and void environments.

Two sub-samples are defined based on the bulge-to-total mass ratio, following \cite{guo2011dwarf}: pure-disk galaxies with ($M_{\text{bulge}}/M_{\text{Total}} < 0.03$) and bulge-dominated galaxies with ($M_{\text{bulge}}/M_{\text{Total}} > 0.7$). The final sample includes 939, 1472, and 1272 pure-disk galaxies in void, group, and cluster environments, respectively, and 125, 199, and 122 bulge-dominated galaxies in the same environments.

As shown in Figure~\ref{Fig:2}, satellite abundance increases with host stellar mass for both pure-disk (left panel) and bulge-dominated (right panel) systems across all environments. A clear environmental dependence is evident: at fixed stellar mass, cluster hosts consistently contain fewer satellites than group and void hosts, irrespective of the morphological type.

To quantitatively assess the apparent differences at the high-mass end visible in Figure~\ref{Fig:2}, we performed non-parametric Mann-Whitney U tests (\cite{Mann1947}; see also \cite{Kumamoto2019} for an application in astrophysics) for each mass bin (width 0.5 dex) and environment separately. Table~\ref{tab:pvalues} reports the resulting p-values. In voids, the number of high-mass galaxies in the two morphological classes is insufficient for a robust test (marked as `non'). In clusters, all p-values exceed 0.05, indicating no statistically significant difference between pure-disk and bulge-dominated galaxies across the entire mass range. In groups, however, for the highest two mass bins ($10.0$–$10.5$ and $10.5$–$11.0$), we obtain p-values of $0.009$ and $<0.001$, respectively. These values reveal a significant difference between the two morphological types in group environments at high stellar masses: bulge-dominated group galaxies host systematically more satellites than pure-disk galaxies of the same stellar mass. This finding is consistent with \cite{Muller2023}, who showed that early-type galaxies (which typically have large bulges) possess richer satellite populations. The lack of morphological dependence in clusters suggests that dense-environment processes (e.g. ram-pressure stripping, harassment) efficiently suppress satellite populations regardless of host morphology. In contrast, the significant morphological dependence in groups at high masses implies that in less extreme environments, the host's internal structure (specifically bulge growth) plays a role in shaping satellite abundance. These results indicate that the influence of host morphology on satellite populations is environment-dependent, detectable only in groups and at high stellar masses, while clusters erase this signal through efficient environmental quenching.

\begin{table}[h]
\centering
\small
\caption{p-values from the Mann-Whitney U test comparing pure-disk vs bulge-dominated satellite counts per mass bin and environment.}
\label{tab:pvalues}
\begin{tabular}{lccc}
\hline
Mass bin ($\log_{10}M_*/M_{\odot}$) & Void & Group & Cluster \\
\hline
8.0 -- 8.5 & 0.105 & 0.121 & 0.340 \\
8.5 -- 9.0 & 0.082 & 0.034 & 0.209 \\
9.0 -- 9.5 & 0.377 & 0.381 & 0.560 \\
9.5 -- 10.0 & non & 0.607 & 0.631 \\
10.0 -- 10.5 & non & 0.009 & 0.251 \\
10.5 -- 11.0 & non & $<$0.001 & 0.288 \\
\hline
\end{tabular}
\par\smallskip\noindent Note. 'non' indicates fewer than three galaxies in at least one morphological sample, preventing a reliable test.
\end{table}

\subsubsection{Dependence on specific star formation rate and stellar disk scale length
    \label{sec:3.1.3}}

\begin{figure}[h!]
    \centering
    \includegraphics[width=0.33\textwidth]{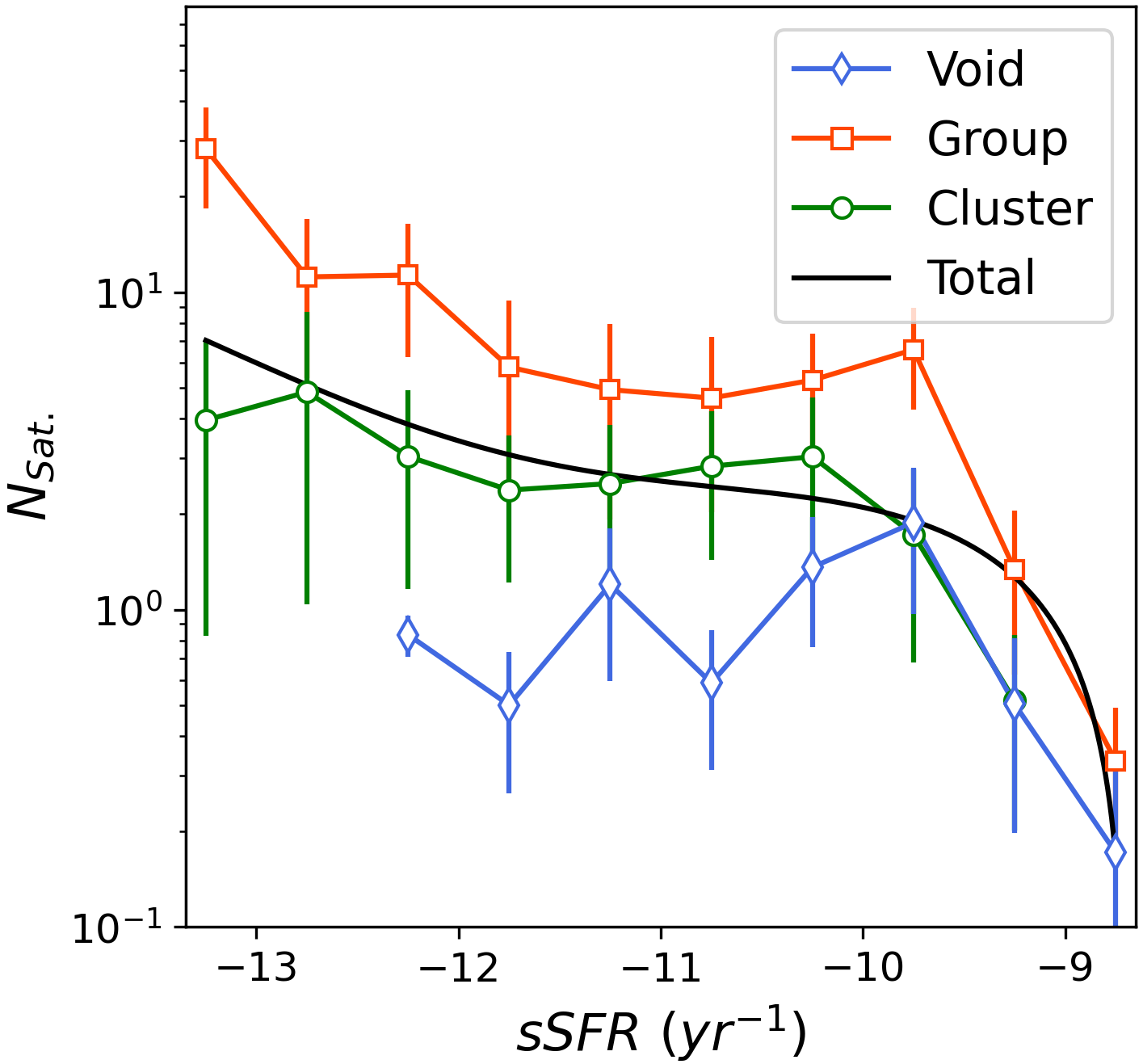}
    \caption{Mean satellite number ($N_{\text{sat}}$) as a function of specific star formation rate (sSFR). Different environments are indicated by colours: voids (blue), groups (orange), clusters (green), and the full population (black). Error bars represent $1\sigma$ uncertainties.}
    \label{Fig:3}
\end{figure}

The relationship between the specific star formation rate (sSFR) and satellite abundance reveals pronounced environmental dependencies. As shown in Figure~\ref{Fig:3}, group and cluster environments exhibit a significant anti-correlation: quiescent host galaxies with lower sSFR systematically host richer satellite systems than their star-forming counterparts. This trend intensifies with decreasing sSFR, indicating that the cessation of star formation in central galaxies is closely associated with the buildup of more abundant satellite populations. The clear monotonic relation suggests that satellite abundance may serve as a reliable tracer of the host galaxy evolutionary stage in dense environments.

In contrast, void environments display fundamentally different behavior, showing no significant correlation between sSFR and satellite abundance. Across the limited dynamic range of sSFR values in these underdense regions, where galaxies are predominantly star-forming, the number of satellites remains approximately constant. This absence of correlation suggests that the mechanisms connecting star formation quenching and satellite system assembly in dense environments are either absent or substantially weakened in voids.

\begin{figure*}[h!]
    \sidecaption
    \includegraphics[width=0.7\hsize]{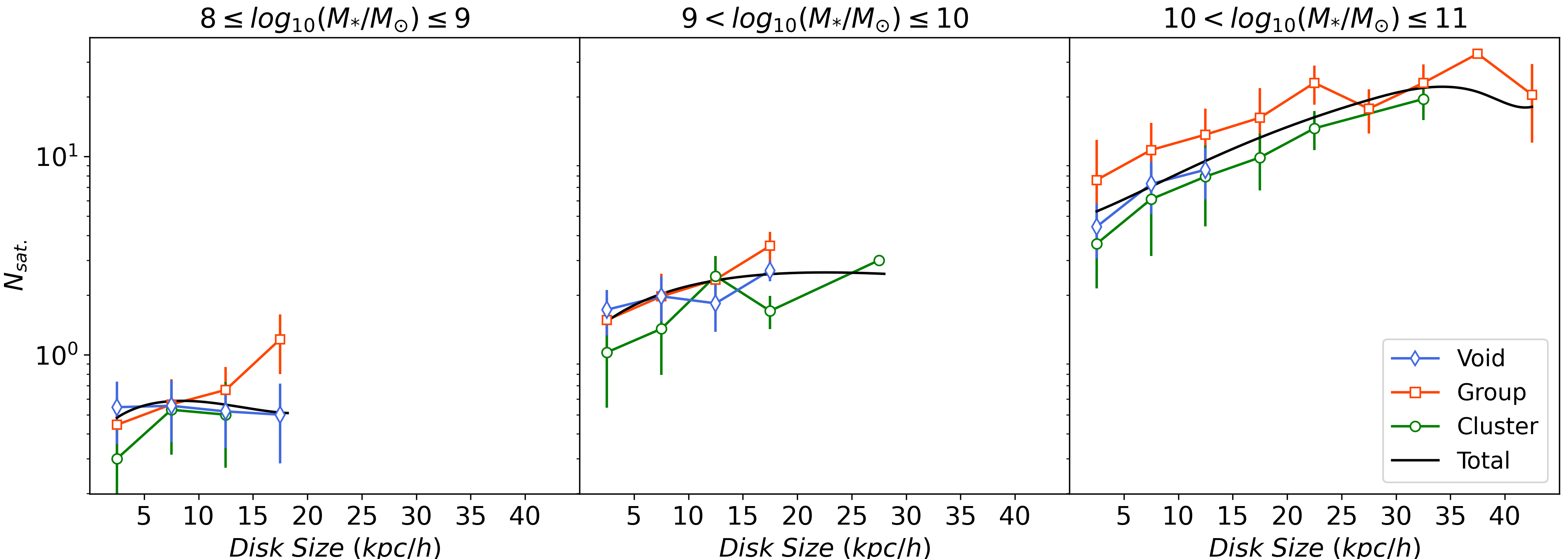}
    \caption{Mean satellite number ($N_{\text{sat}}$) as a function of stellar disk scale length, shown for three stellar mass bins: left: $8<\log_{10}(M_*/M_{\odot})<9$, middle: $9<\log_{10}(M_*/M_{\odot})<10$, right: $10<\log_{10}(M_*/M_{\odot})<11$. Colours indicate environments: voids (blue), groups (orange), clusters (green), and the full population (black). Error bars represent $1\sigma$ uncertainties.}
    \label{Fig:4}
\end{figure*}

Because satellite abundance is known to correlate strongly with host stellar mass, we first divided the sample into three stellar mass bins to isolate the effect of disk scale length: $8<\log_{10}(M_*/M_{\odot})<9$, $9<\log_{10}(M_*/M_{\odot})<10$, and $10<\log_{10}(M_*/M_{\odot})<11$. Figure~\ref{Fig:4} presents the mean satellite number as a function of disk scale length for voids (blue), groups (orange), and clusters (green) in each mass bin. In the lowest mass bin (left panel), void galaxies show a flat relation: satellite abundance remains constant with disk size. Group galaxies exhibit a very weak positive trend, but the slope is shallow. In the intermediate mass bin (middle panel), all three environments display a positive correlation, albeit with considerable scatter. In the highest mass bin (right panel), a strong positive correlation is evident in all environments: larger disks host more satellites. Strikingly, group galaxies dominate the satellite population across the entire disk size range, hosting significantly more satellites than both void and cluster galaxies. This enhancement in groups is robust and becomes increasingly pronounced at higher masses. We conclude that the environmental dependence of the disk size–satellite relation is genuine and mass-dependent. The particularly rich satellite systems of massive group galaxies with large disks suggest that groups provide an optimal environment for satellite accretion and survival, balancing sufficient gravitational potential with reduced disruptive processes compared to clusters.

The environmental dependence of both relationships provides valuable constraints on galaxy formation models. The strong anti-correlation with sSFR in dense environments supports scenarios in which environmental processes (e.g. strangulation, \cite{peng2015strangulation}; ram-pressure stripping, \cite{boselli2022ram}; and galaxy harassment, \cite{bialas2015occurrence}) simultaneously quench star formation in central galaxies and shape their satellite populations through ongoing accretion. Likewise, the positive correlation between disk size and satellite abundance in groups and clusters suggests that disk growth and satellite system assembly are influenced by similar processes, including mergers and accretion.

\subsection{Environmental and cosmic evolution of satellite systems 
	\label{sec:3.2}}
This section examines the evolution of satellite systems as a function of both a large-scale environment and cosmic time. Analysis of radial profiles and abundances across different host mass bins reveals systematic trends that reflect the combined influence of environment and redshift on the assembly and evolution of satellite populations.

\subsubsection{Environmental trends in radial profiles
    \label{sec:3.2.1}}

To investigate the spatial distribution of satellite galaxies relative to their hosts, radial profiles were computed by normalizing satellite distances to the virial radius of each central galaxy. A stacking method was then applied, combining the profiles of all galaxies within each environment (clusters, groups, and voids) and normalizing the cumulative satellite counts in each radial bin by the total number of satellites in that environment. This approach enables a direct comparison of relative satellite distributions across environments. The analysis was carried out in three stellar mass bins: $8 < \log(M_*/M_{\odot}) < 9$, $9 < \log(M_*/M_{\odot}) < 10$, and $10 < \log(M_*/M_{\odot}) < 11$, separately for three morphological classes: all morphologies, pure disk, and bulge dominated (see Section~\ref{sec:3.1.2} for definitions). All results presented here are based on the $z=0$ snapshot of the simulation (Section~\ref{sec:Mill}).

\begin{figure*}[h!]
	\sidecaption
	\includegraphics[width=0.7\hsize]{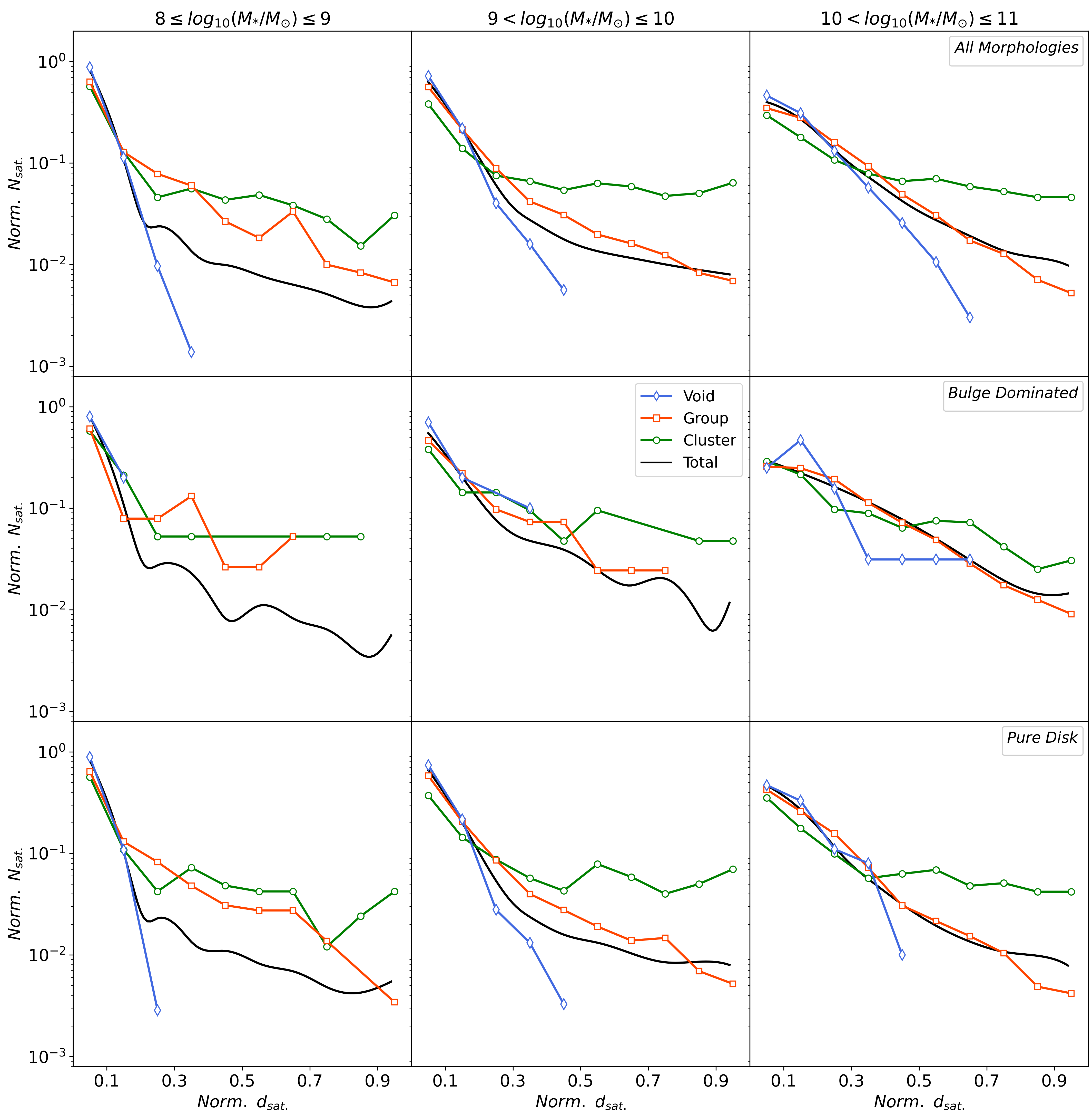}	
\caption{Normalized radial profiles of satellite number as a function of normalized distance to the host galaxy, shown for three morphological classes (all morphologies, bulge dominated, pure disk) and three stellar mass bins (low, intermediate, high). Different environments are depicted using distinct colours.}
	\label{Fig:5}
\end{figure*}

As shown in Figure~\ref{Fig:5}, the resulting profiles reveal systematic environmental trends that persist across all mass bins. In this figure, cluster profiles are shown in green, group profiles in orange, void profiles in blue, and the full galaxy sample in black.

For all morphologies, void environments exhibit a markedly stronger central concentration, with a larger fraction of satellites residing close to the host galaxy centre. This enhanced concentration suggests that satellite systems in voids experience relatively little environmental disruption, thereby preserving their initial radial distributions. In contrast, cluster environments display substantially flatter profiles. At larger radii, the relative abundance of satellites in clusters exceeds that in both groups and voids. This flattening indicates that cluster-specific processes (e.g. tidal stripping, galaxy harassment, and dynamical friction) redistribute satellites towards larger radii or preferentially disrupt those in the inner regions. Group environments consistently show intermediate behaviour, with radial profiles lying between those of clusters and voids across all stellar mass ranges. This systematic progression — from centrally concentrated void profiles through intermediate group profiles to flattened cluster profiles — demonstrates a continuous environmental influence on satellite spatial distributions.

For bulge-dominated galaxies in the lowest mass bin ($8-9$), the radial profiles show considerable scatter due to the limited number of objects. Nevertheless, void environments consistently exhibit a more concentrated satellite distribution compared to groups and clusters. In clusters and groups, the profiles are flatter, with satellite abundance becoming more prominent at larger radii. As we move to intermediate ($9-10$) and high ($10-11$) mass bins, the central concentration in the voids becomes less pronounced than in the low‑mass bin, but the voids still retain a more concentrated profile than clusters and groups. Cluster profiles remain the flattest across all mass bins, and the excess of satellites at large radii is clearly visible in clusters, especially at high masses.

For pure-disk galaxies, the radial profiles follow the same environmental trends as pure‑-disk hosts in all three mass bins. However, a notable difference is the much steeper contrast between clusters and the other environments: in clusters, the flattening of the profile is more pronounced, and the excess of satellites at large radii is significantly stronger than in pure‑-disk systems. This difference is evident across the entire mass range and becomes more prominent at higher masses.

Importantly, these environmental trends are robust across all three stellar mass bins, implying that the large‑scale environment is the primary driver of satellite spatial redistribution. The consistency of these patterns suggests that environmental mechanisms act in a broadly similar manner across different host mass scales, though with varying efficiency. The concentrated distributions in voids likely reflect relatively undisturbed systems, where satellites retain their initial radial distribution. The flatter profiles in groups, and even more so in clusters, point to efficient environmental processing. In particular, the stronger flattening and the enhanced outer‑-radius satellite excess in bulge‑dominated cluster hosts compared to pure‑disk ones may be explained by their older assembly histories and deeper potential wells, which make them more susceptible to tidal stripping and harassment over cosmic time. Conversely, less pronounced flattening in pure-disk cluster hosts suggests that their shallower potential wells and younger dynamical ages result in a milder modification of their satellite radial profiles. Overall, the radial profiles demonstrate that environmental mechanisms (tidal stripping, dynamical friction, and harassment) preferentially affect inner satellites or redistribute them outward, with an efficiency that depends on both the environment and the host morphology.

\subsubsection{Redshift evolution of satellite radial profiles 
	\label{sec:3.2.2}}

\begin{figure*}[h!]
	\sidecaption
	\includegraphics[width=0.7\hsize]{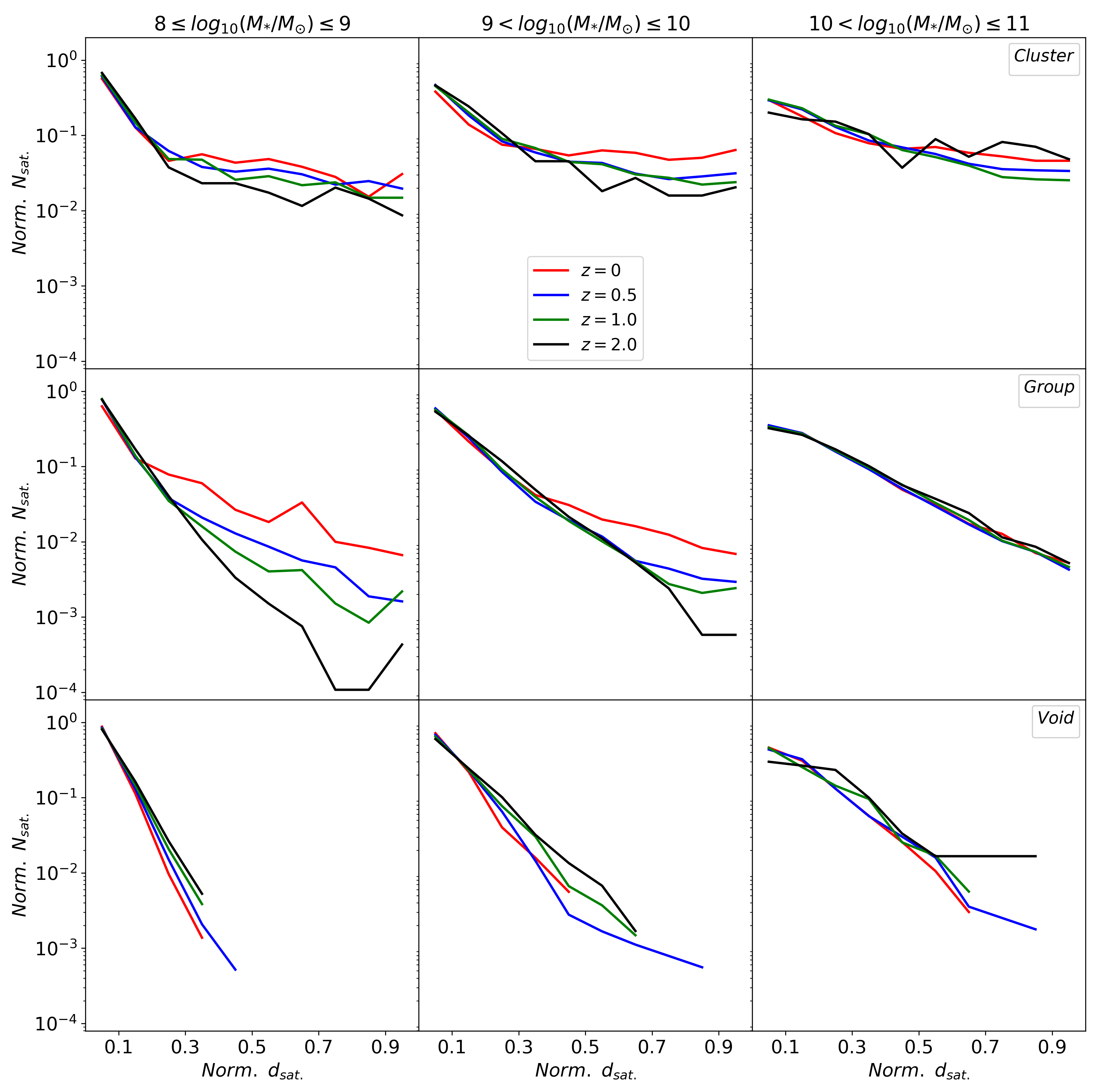}
	\caption{Evolution of normalized radial profiles of satellite number ($N_{\text{sat}}$) as a function of normalized distance to the host galaxy ($d_{\text{sat}}/R_{\text{vir}}$). Columns correspond to stellar mass bins: low-mass (left), intermediate-mass (middle), and high-mass (right). Rows correspond to environments: clusters (top), groups (middle), and voids (bottom).}
	\label{Fig:6}
\end{figure*}

The evolutionary pathways of satellite radial profiles were investigated through their redshift dependence across different environments and host mass regimes. Figure \ref{Fig:6} illustrates how the profiles vary systematically with both environment and cosmic epoch, enabling a direct comparison of evolutionary trends between clusters, groups, and voids. The analysis was carried out in three host stellar mass bins ($8 < \log(M_*/M_\odot) < 9$, $9 < \log(M_*/M_\odot) < 10$, and $10 < \log(M_*/M_\odot) < 11$), with profiles shown at four redshifts: $z=0$ (red), $z=0.5$ (blue), $z=1$ (green), and $z=2$ (black).

In cluster and group environments, low- and intermediate-mass hosts ($8 < \log(M_*/M_\odot) < 10$) display clear evolutionary trends. As the redshift decreases from $z=2$ to $z=0$, the radial profiles become progressively less concentrated, with satellites distributed more broadly around their hosts. This flattening indicates that dense environment processes such as tidal stripping (\cite{hahn2009tidal}), dynamical friction (\cite{fujii2006dynamical}), and galaxy harassment (\cite{bialas2015occurrence}), redistribute satellites toward larger radii or preferentially deplete the inner regions over time.

By contrast, high-mass hosts ($10 < \log(M_*/M_\odot) < 11$) in clusters and groups show remarkable stability. Their radial profiles remain nearly unchanged across the full redshift range, suggesting that satellite systems around massive centrals in dense environments reach a dynamically evolved configuration early, with subsequent evolution producing only minor modifications.

Void environments reveal the opposite evolutionary behaviour. Across all mass bins, satellites are consistently more centrally concentrated than in dense environments, and this concentration increases further towards a lower redshift. This trend implies that in the absence of disruptive external mechanisms, satellite systems in voids evolve primarily through internal processes such as dynamical friction and mergers, leading to a gradual buildup of satellites near the host centre.

The contrast between dense and underdense environments underscores the decisive role of the large-scale environment in shaping satellite system evolution. While clusters and groups drive satellites outwards and flatten radial profiles over time, voids promote increasing central concentration through quiescent internal pathways. These environment-driven evolutionary patterns persist across all stellar mass bins, though with varying amplitude. The stability of profiles around massive hosts in dense environments further suggests that the efficiency of environmental processing depends jointly on host mass and environmental density.

\subsubsection{Cosmic evolution of satellite abundance 
	\label{sec:3.2.3}}
\begin{figure*}[h!]
	\centering
	\sidecaption
	\includegraphics[width=0.7\hsize]{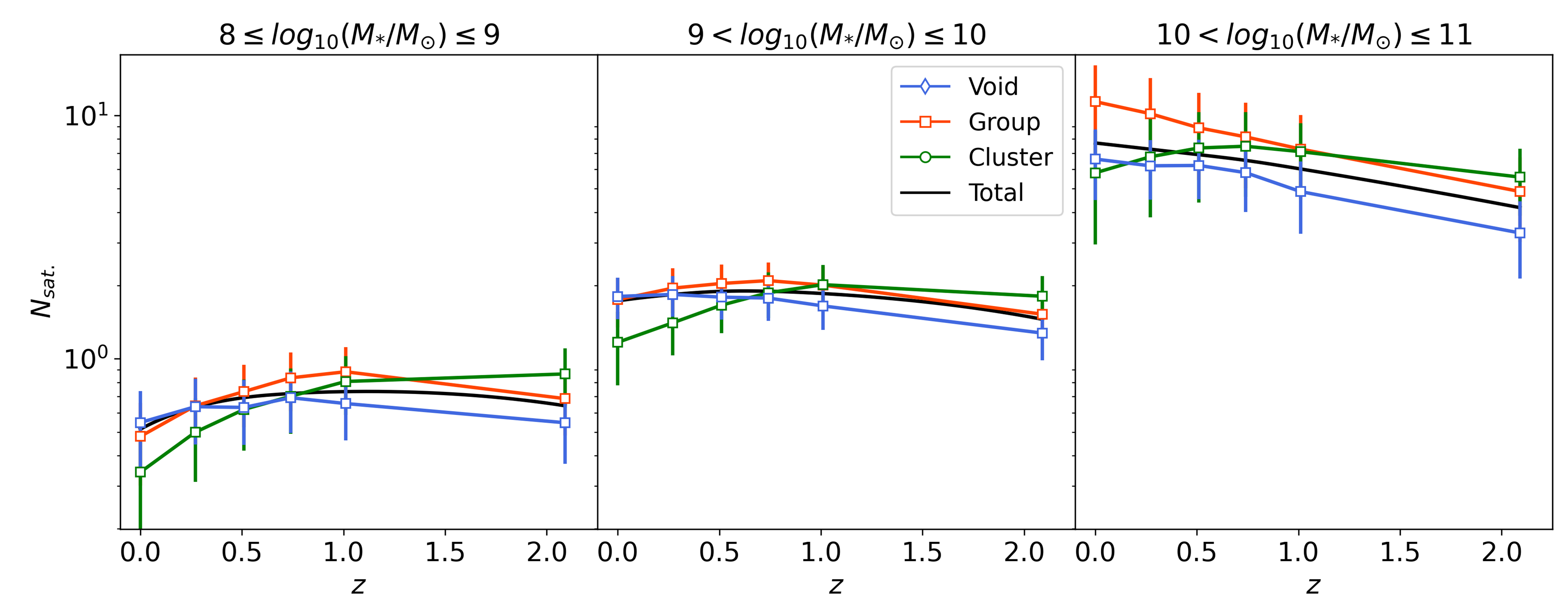}
	\caption{Mean trend of satellite number ($N_{sat.}$) as a function of redshift ($z$) is shown for low-, intermediate-, and high-mass host galaxies in left, middle and right panels, respectively. Different environments are illustrated using distinct colours.}
	\label{Fig:7}
\end{figure*}

The evolutionary history of satellite systems was traced by following host galaxies from redshift $z=2$ to $z=0$ across different environments and stellar mass regimes. Figure \ref{Fig:7} shows the evolution of the mean satellite abundance for hosts in three stellar mass bins (low-mass: $8 < \log(M_*/M_\odot) < 9$, intermediate-mass: $9 < \log(M_*/M_\odot) < 10$, and high-mass: $10 < \log(M_*/M_\odot) < 11$) within cluster (green), group (orange), and void (blue) environments.

In voids, the satellite abundance gradually increases and monotonically from $z=2$ to $z=0$ across all mass ranges. This slow accumulation suggests steady growth through continuous accretion and in situ formation, largely unaffected by disruptive environmental processes.

Groups display more complex, mass-dependent trends. For low- and intermediate-mass hosts, satellite abundance increases from $z=2$ to $z=1$ and then declines towards $z=0$, indicating a transition from net accretion to disruption or mergers at later epochs (\cite{martin2021role}). In contrast, high-mass hosts show a continuous increase from $z=2$ to $z=0$, implying that massive group hosts continue to accumulate satellites efficiently throughout cosmic time.

Clusters exhibit the most dramatic evolution. Across all mass bins the satellite abundance remains nearly constant or increases slightly from $z=2$ to $z=1$, followed by a sharp decline towards $z=0$. This pronounced decrease points to intense environmental processing during the last $\sim$8 Gyr, where tidal disruption, harassment, and mergers strongly reduce satellite populations. The suppression is far more severe in clusters than in groups or voids.

These contrasting evolutionary pathways highlight the decisive role of the environment in shaping satellite system assembly.  Although voids enable gradual accumulation, clusters drive strong late-time depletion, and groups occupy an intermediate regime where outcomes depend sensitively on host mass. These results demonstrate that environmental processes determine not only present-day satellite abundances but also govern their full assembly histories across cosmic time.

\section{Summary and conclusion \label{sec:sum}}

This study investigated the environmental dependence of dwarf satellite galaxy populations using the Millennium-II simulation combined with the semi-analytic model of \cite{guo2011dwarf} (\ref{sec:Mill}). Consistent samples of host galaxies and their satellites were constructed across cluster, group, and void environments (\ref{sec:environment}) to quantify how satellite abundance and spatial distribution scale with host properties and vary across the cosmic density spectrum (\ref{sec:satellite}).

Satellite abundance is found to be primarily governed by the stellar and bulge mass of the host (\ref{sec:3.1.1}), with cluster environments consistently suppressing satellite populations relative to groups and voids through mechanisms such as tidal and ram-pressure stripping. Interestingly, at fixed total stellar mass, groups and voids host similar numbers of satellites, but at fixed bulge mass, groups exhibit a significant excess over voids. This demonstrates that the enhancement of satellite populations around bulge-dominated galaxies in groups is a genuine environmental effect, not merely a reflection of stellar mass. Once the stellar mass is accounted for, host morphology (\ref{sec:3.1.2}) exerts minimal additional influence, underscoring the dominant role of the gravitational potential. Correlations with specific star formation rate and disk scale length (\ref{sec:3.1.3}) arise only in dense environments, indicating that group and cluster conditions link host galaxy evolution to the assembly of satellite systems, whereas such connections are absent in voids.

At $z=0$, the radial profiles (\ref{sec:3.2.1}) confirm that the environment dominates the host mass in shaping satellite distributions: the voids show strong central concentrations, the clusters exhibit flattened profiles due to intense processing, and the groups display intermediate behaviour. These trends are consistently observed for both pure‑disk and bulge‑dominated hosts. Notably, in clusters, bulge‑dominated galaxies show a stronger flattening and a greater excess of satellites at large radii than pure-disk systems, pointing to more efficient tidal processing of older, deeper potential wells. These robust patterns across all mass bins and morphological classes emphasize the universal influence of environment on satellite system evolution. Their redshift dependence (\ref{sec:3.2.2}) further reveals the interplay between environment, host mass, and cosmic time: low- and intermediate-mass hosts in clusters and groups show progressive flattening from $z=2$ to $z=0$, while massive hosts remain nearly stable, suggesting early dynamical maturity. In voids, satellites become increasingly concentrated toward lower redshift, consistent with quiescent internal growth.

The cosmic evolution of satellite abundance (\ref{sec:3.2.3}) highlights distinct environment-dependent pathways. Voids foster gradual accumulation, groups show mass-dependent trends with late-time depletion for lighter hosts but continued growth for massive ones, and clusters undergo the strongest suppression after $z=1$. These contrasting patterns demonstrate that the environment is the primary driver of satellite system evolution, with outcomes modulated by host mass and cosmic epoch.

Overall, the results underscore the decisive role of cosmic structures and galactic environments in governing the distribution and abundance of satellite galaxies. From the dense interiors of clusters to the relative isolation of voids, the large-scale environment imprints systematic signatures on the assembly and evolution of satellite systems. Thus, satellite populations cannot be fully characterized based solely on host galaxy properties; rather, they must be interpreted within the broader cosmological context, wherein the interplay between structure formation and environmental processes governs their abundance and spatial configuration over time.

Future wide-field surveys will offer unprecedented opportunities to test the predictions presented in this work. Deep imaging from the Vera C. Rubin Observatory’s LSST (\cite{thomas2020vera}) will significantly expand the census of faint dwarf satellites around Milky Way-like hosts, while space-based missions such as Euclid (\cite{racca2016euclid}) and the Nancy Grace Roman Space Telescope (\cite{wang2022high}) will systematically map satellite systems across diverse environments with uniform depth and coverage. Moreover, observational data on satellite populations at higher redshifts will provide critical tests of their evolutionary dependence on host galaxy environments.

In parallel, cosmological simulations incorporating diverse gravo-magnetohydrodynamic frameworks (e.g. IllustrisTNG (\cite{nelson2019illustristng})) and semi-analytical models, updated with cosmological parameters from Plank-2016 (\cite{ade2016planck}) rather than the WMAP-based values adopted in this work, will enable stringent validation of these results.

In addition, applying a range of cosmic‑web classification techniques including alternative group and cluster finders (e.g. \cite{graham2023group}), void identification methods (e.g. \cite{Ghafourvega2025}), and complementary approaches that incorporate other environments such as filaments and walls (e.g. \cite{Ghafourgravipast2025,cautun2013nexus}), would allow these environmental trends to be examined in greater depth. Such an analysis would further refine our understanding of how host properties and the surrounding large‑scale structure jointly influence the assembly of satellite systems across cosmic time.

\bibliographystyle{aa}
\bibliography{References}

\end{document}